\documentclass[11pt,prd,letterpaper,amsmath,amssymb,final,nofootinbib
,unsortedaddress,superscriptaddress
]{revtex4-1}

\pdfoutput=1

\usepackage[top=0.9in, bottom=0.9in, left = 0.9in, right=0.9in]{geometry}

\usepackage{graphicx}
\usepackage{dcolumn}
\usepackage{bbm}
\usepackage{textcomp}
\usepackage{color}
\usepackage{gensymb}
\usepackage{wasysym}
\usepackage{amssymb}
\usepackage{enumerate}
\usepackage{geometry}
\usepackage{setspace}
\usepackage{comment}
\usepackage[utf8]{inputenc}	
\usepackage{mathrsfs}
\usepackage{subcaption}
\usepackage[T1]{fontenc}
\usepackage{amssymb,amsmath}
\usepackage{graphicx, amsthm}
\usepackage{booktabs}
\usepackage{isotope}
\usepackage[graphicx]{realboxes}
\usepackage{xspace}

\newcommand{\theia}[0]{\textsc{Theia}\xspace}
  
\begin{document}

\title{\theia: Summary of physics program\\ {\small Snowmass White Paper Submission}}

\author{M.~Askins}\affiliation{\ucb}\affiliation{\lbnl}
\author{Z.~Bagdasarian}\affiliation{\ucb}\affiliation{\lbnl}
\author{N.~Barros}\affiliation{\penn}
\affiliation{\fcul}
\affiliation{ \lip}
\author{E.W.~Beier}\affiliation{\penn}
\author{A.~Bernstein}\affiliation{\llnl}
\author{M.~B\"ohles}\affiliation{\mainz}
\author{E.~Blucher}\affiliation{\chic}
\author{R.~Bonventre}\affiliation{\lbnl}
\author{E.~Bourret}\affiliation{\lbnl}
\author{E.~J.~Callaghan}\affiliation{\ucb}\affiliation{\lbnl}
\author{J.~Caravaca}\affiliation{\ucb}\affiliation{\lbnl}
\author{M.~Diwan}\affiliation{\bnl}
\author{S.T.~Dye}\affiliation{\uh}
\author{J.~Eisch}\affiliation{\fnal}
\author{A.~Elagin}\affiliation{\chic}
\author{T.~Enqvist}\affiliation{\jyv}
\author{U.~Fahrendholz}\affiliation{\mun}
\author{V.~Fischer}\affiliation{\ucd}
\author{K.~Frankiewicz}\affiliation{\bu}
\author{C.~Grant}\affiliation{\bu}
\author{D.~Guffanti}\affiliation{\mainz}
\author{C.~Hagner}\affiliation{\ham}
\author{A.~Hallin}\affiliation{\alb}
\author{C.~M.~Jackson}\affiliation{\pnnl}
\author{R.~Jiang}\affiliation{\chic}
\author{T.~Kaptanoglu}\affiliation{\ucb}\affiliation{\lbnl}
\author{J.R.~Klein}\affiliation{\penn}
\author{Yu.~G.~Kolomensky}\affiliation{\ucb}\affiliation{\lbnl}
\author{C.~Kraus}\affiliation{\laur}
\author{F.~Krennrich}\affiliation{\iowa}
\author{T.~Kutter}\affiliation{\lsu}
\author{T.~Lachenmaier}\affiliation{\tub}
\author{B.~Land}\affiliation{\ucb}\affiliation{\lbnl}\affiliation{\penn}
\author{K.~Lande}\affiliation{\penn}
\author{L.~Lebanowski}\affiliation{\penn}
\author{J.G.~Learned}\affiliation{\uh}
\author{V.A.~Li}\affiliation{\llnl}
\author{V.~Lozza}\affiliation{\fcul}
\affiliation{ \lip}
\author{L.~Ludhova}\affiliation{\jul}
\author{M.~Malek}\affiliation{\sheff}
\author{S.~Manecki}\affiliation{\laur}\affiliation{\qu}\affiliation{\snolab}
\author{J.~Maneira}\affiliation{\fcul}
\affiliation{ \lip}
\author{J.~Maricic}\affiliation{\uh}
\author{J.~Martyn}\affiliation{\mainz}
\author{A.~Mastbaum}\affiliation{\rut}
\author{C.~Mauger}\affiliation{\penn}
\author{M.~Mayer}\affiliation{\mun}
\author{J.~Migenda}\affiliation{\kings}
\author{F.~Moretti}\affiliation{\lbnl}
\author{J.~Napolitano}\affiliation{\temp}
\author{B.~Naranjo}\affiliation{\ucla}
\author{S.~Naugle}\affiliation{\penn}
\author{M.~Nieslony}\affiliation{\mainz}
\author{L.~Oberauer}\affiliation{\mun}
\author{G.~D.~Orebi~Gann}\affiliation{\ucb}\affiliation{\lbnl}
\author{J.~Ouellet}\affiliation{\mitnew}
\author{T.~Pershing}\affiliation{\ucd}
\author{S.T.~Petcov}\affiliation{\tri,\kav}
\author{L.~Pickard}\affiliation{\ucd}
\author{R.~Rosero}\affiliation{\bnl}
\author{M.~C.~Sanchez}\affiliation{\iowa}
\author{J.~Sawatzki}\affiliation{\mun}
\author{S.~Schoppmann}\affiliation{\ucb}\affiliation{\lbnl}
\author{S.H.~Seo}\affiliation{\kor}
\author{M.~Smiley}\affiliation{\ucb}\affiliation{\lbnl}
\author{M.~Smy}\affiliation{\uci}
\author{A.~Stahl}\affiliation{\aach}
\author{H.~Steiger}\affiliation{\mainz}\affiliation{\mun}
\author{M.~R.~Stock}\affiliation{\mun}
\author{H.~Sunej}\affiliation{\bnl}
\author{R.~Svoboda}\affiliation{\ucd}
\author{E.~Tiras}\affiliation{\erc}\affiliation{\iow}
\author{W.~H.~Trzaska}\affiliation{\jyv}
\author{M.~Tzanov}\affiliation{\lsu}
\author{M.~Vagins}\affiliation{\uci}
\author{C.~Vilela}\affiliation{\sbu}
\author{Z.~Wang}\affiliation{\tsing}
\author{J.~Wang}\affiliation{\ucd}
\author{M.~Wetstein}\affiliation{\iowa}
\author{M.J.~Wilking}\affiliation{\sbu}
\author{L.~Winslow}\affiliation{\mitnew}
\author{P.~Wittich}\affiliation{\corn}
\author{B.~Wonsak}\affiliation{\ham}
\author{E.~Worcester}\affiliation{\bnl}\affiliation{\sbu}
\author{M.~Wurm}\affiliation{\mainz}
\author{G.~Yang}\affiliation{\sbu}
\author{M.~Yeh}\affiliation{\bnl}
\author{E.D.~Zimmerman}\affiliation{\boul}
\author{S.~Zsoldos}\affiliation{\ucb}\affiliation{\lbnl}
\author{K.~Zuber}\affiliation{\dres}

\newcommand{\ucb}{Physics Department, University of California at Berkeley, Berkeley, CA 94720-7300
}
\newcommand{\lbnl}{
Lawrence Berkeley National Laboratory, 1 Cyclotron Road, Berkeley, CA 94720-8153, USA
}
\newcommand{\jul}{Forschungszentrum J{\"u}lich, Institute for Nuclear Physics, Wilhelm-Johnen-Stra{\ss}e 52425 J{\"u}lich, Germany
}\newcommand{\penn}{Department of Physics and Astronomy, University of Pennsylvania, Philadelphia, PA 19104-6396
}\newcommand{\fcul}{Universidade de Lisboa, Faculdade de Ci{\^e}ncias (FCUL), Departamento de F{\'i}sica, Campo Grande, Edifício C8, 1749-016 Lisboa, Portugal
}\newcommand{\lip}{Laborat{\'o}rio de Instrumenta{}{\c c}{\~a}o e F{\'i}sica Experimental de Part{\'i}culas (LIP), Av. Prof. Gama Pinto, 2, 1649-003, Lisboa, Portugal
}\newcommand{\chic}{
The Enrico Fermi Institute and Department of Physics, The University of Chicago, Chicago, IL 60637, USA
}\newcommand{\bnl}{
Brookhaven National Laboratory, Upton, New York 11973, USA
}\newcommand{\uh}{
University of Hawai‘i at Manoa, Honolulu, Hawai‘i 96822, USA
}\newcommand{\iowa}{
Department of Physics and Astronomy, Iowa State University, Ames, IA 50011, USA
}\newcommand{\jyv}{
Department of Physics, University of Jyv{\"a}skyl{\"a}, Finland
}\newcommand{\ucd}{
University of California, Davis, 1 Shields Avenue, Davis, CA 95616, USA
}\newcommand{\bu}{
Boston University, Department of Physics, Boston, MA 02215, USA
}\newcommand{\mainz}{
Institute of Physics and Excellence Cluster PRISMA, Johannes Gutenberg-Universit{\"a}t Mainz, 55099 Mainz, Germany
}\newcommand{\ham}{
Institut f{\"u}r Experimentalphysik, Universit{\"a}t Hamburg, 22761 Hamburg, Germany
}\newcommand{\alb}{
University of Alberta, Department of Physics, 4-181 CCIS, Edmonton, AB T6G 2E1, Canada
}\newcommand{\pnnl}{
Pacific Northwest National Laboratory, Richland, WA 99352, USA
}\newcommand{\laur}{
Laurentian University, Department of Physics, 935 Ramsey Lake Road, Sudbury, ON P3E 2C6, Canada
}\newcommand{\lsu}{
Department of Physics and Astronomy, Louisiana State University, Baton Rouge, LA 70803
}\newcommand{\tub}{
Kepler Center for Astro and Particle Physics, Universit{\"a}t T{\"u}bingen, 72076 T{\"u}bingen, Germany
}\newcommand{\sheff}{
University of Sheffield, Physics \& Astronomy, Western Bank, Sheffield S10 2TN, UK
}\newcommand{\qu}{
Queen's University, Department of Physics, Engineering Physics \& Astronomy, Kingston, ON K7L 3N6, Canada
}\newcommand{\snolab}{
SNOLAB, Creighton Mine 9, 1039 Regional Road 24, Sudbury, ON P3Y 1N2, Canada
}\newcommand{\rut}{
Department of Physics and Astronomy, Rutgers, The State University of New Jersey, 136 Frelinghuysen Road, Piscataway, NJ 08854-8019 USA
}\newcommand{\temp}{
Department of Physics, Temple University, Philadelphia, PA, USA
}\newcommand{\ucla}{
University of California Los Angeles, Department of Physics \& Astronomy, 475 Portola Plaza, Los Angeles, CA 90095-1547, USA
}
\newcommand{\tri}{
SISSA/INFN, Via Bonomea 265, I-34136 Trieste, Italy
}\newcommand{\kav}{
Kavli IPMU (WPI), University of Tokyo, 5-1-5 Kashiwanoha, 277-8583 Kashiwa, Japan
}\newcommand{\kor}{
Center for Underground Physics, Institute for Basic Science, Daejeon 34126, Korea
}\newcommand{\uci}{
University of California, Irvine, Department of Physics, CA 92697, Irvine, USA
}\newcommand{\aach}{
Physikzentrum RWTH Aachen, Otto-Blumenthal-Stra{\ss}e, 52074 Aachen, Germany
}\newcommand{\sbu}{
State University of New York at Stony Brook, Department of Physics and Astronomy, Stony Brook, New York, USA
}\newcommand{\tsing}{
Department of Engineering Physics, Tsinghua University, Beijing 100084, China
}\newcommand{\corn}{
Cornell University, Ithaca, NY, USA
}\newcommand{\boul}{
University of Colorado at Boulder, Department of Physics, Boulder, Colorado, USA
}\newcommand{\dres}{
Institut f{\"u}r Kern und Teilchenphysik, TU Dresden, Zellescher Weg 19, 01069, Dresden, Germany
}
\newcommand{\mun}{
Physik-Department and Excellence Cluster Universe, Technische Universit{\"a}t M{\"u}nchen, 85748 Garching, Germany
}\newcommand{\mitnew}{
Massachusetts Institute of Technology, Department of Physics and Laboratory for Nuclear Science, 77 Massachusetts Ave Cambridge, MA 02139, USA
}
\newcommand{\kings}{King’s College London, Department of Physics, Strand Building, Strand, London WC2R 2LS, UK}
\newcommand{\llnl}{
Lawrence Livermore National Laboratory, Livermore, CA 94550, USA
}
\newcommand{\fnal}{
Fermi National Accelerator Laboratory, Batavia, IL 60510, USA
}
\newcommand{\erc}{Department of Physics, Erciyes University, 38030, Kayseri, Turkey
}
\newcommand{\iow}{Department of Physics and Astronomy, The University of Iowa, Iowa City, Iowa, USA}



\maketitle

\section{The \theia experiment}\label{s:intro}

\theia would be a novel, ``hybrid'' optical neutrino detector, with a rich physics program.  This paper is intended to provide a brief overview of the concepts and physics reach of \theia. Full details can be found in the \theia white paper~\cite{theiawp}.

A broad community is pursuing the idea of ``hybrid'' event detection for optical neutrino detectors.  Hybrid detection of particle interactions seeks to combine the unique topology of Cherenkov light with the increased light yield, and sub-Cherenkov threshold sensitivity of scintillation.  The ability to distinguish the two signals is facilitated by new developments in scintillator, photon detection, and readout technology as well as sophisticated analysis methods.  
The Cherenkov light signature offers electron / muon discrimination at high energy, via ring imaging, and sensitivity to particle direction at low energy.  The scintillation yield offers improved energy and vertex resolution and low-threshold particle detection.
The combination also boasts an additional handle on particle identification from the relative intensity of the two signals.

\theia could make use of a number of novel technologies to achieve successful hybrid event detection. 
Water-based liquid scintillator (WbLS)~\cite{WBLS:2011}, or slow scintillators~\cite{slowscin,slowls} can be used to enhance the Cherenkov signal by either reducing or delaying the scintillation component.  The use of angular as well as timing information, with sufficiently fast photon detectors, can offer discrimination between 
Cherenkov and scintillation light for high-energy events even in a standard scintillator like
LAB-PPO~\cite{CHESS2:2017}.  Separation can also be enhanced using very fast photon detectors, such as LAPPDs (Large Area Picosecond Photo-Detectors)~\cite{lappd0,lappd1}, or spectral sorting using dichroic filters~\cite{dichroic}.
A more complete discussion of the relevant technology, including a number of prototype experiments planned or under construction, is provided in the NF10 white paper on ``Future Advances in Photon-Based Neutrino Detectors''.  

\section{\theia physics program}

\theia, named for the Titan Goddess of light, seeks to make world-leading measurements over as broad range of neutrino physics and astrophysics as possible.
The scientific program would include observations of low- and high-energy solar neutrinos, determination of neutrino mass ordering and measurement of the neutrino CP-violating phase $\delta$, observations of diffuse supernova neutrinos and neutrinos from a supernova burst, sensitive searches for nucleon decay and, ultimately, a search for neutrinoless double beta decay (NLDBD), with sensitivity reaching the normal ordering regime of neutrino mass phase space.

\begin{figure*}[htp!]
\centering
\includegraphics[width=0.59\textwidth]{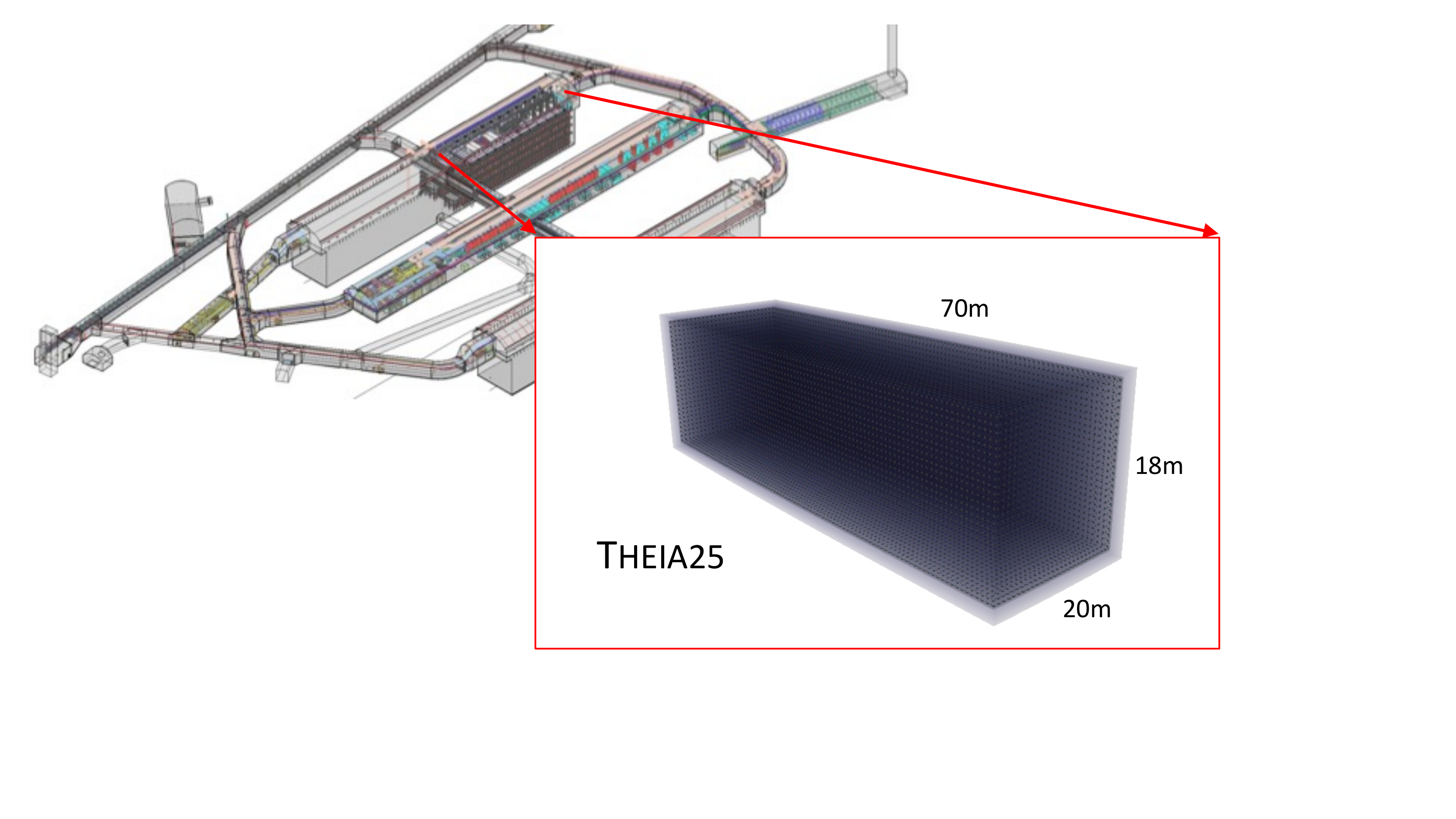}
\includegraphics[width=0.36\textwidth]{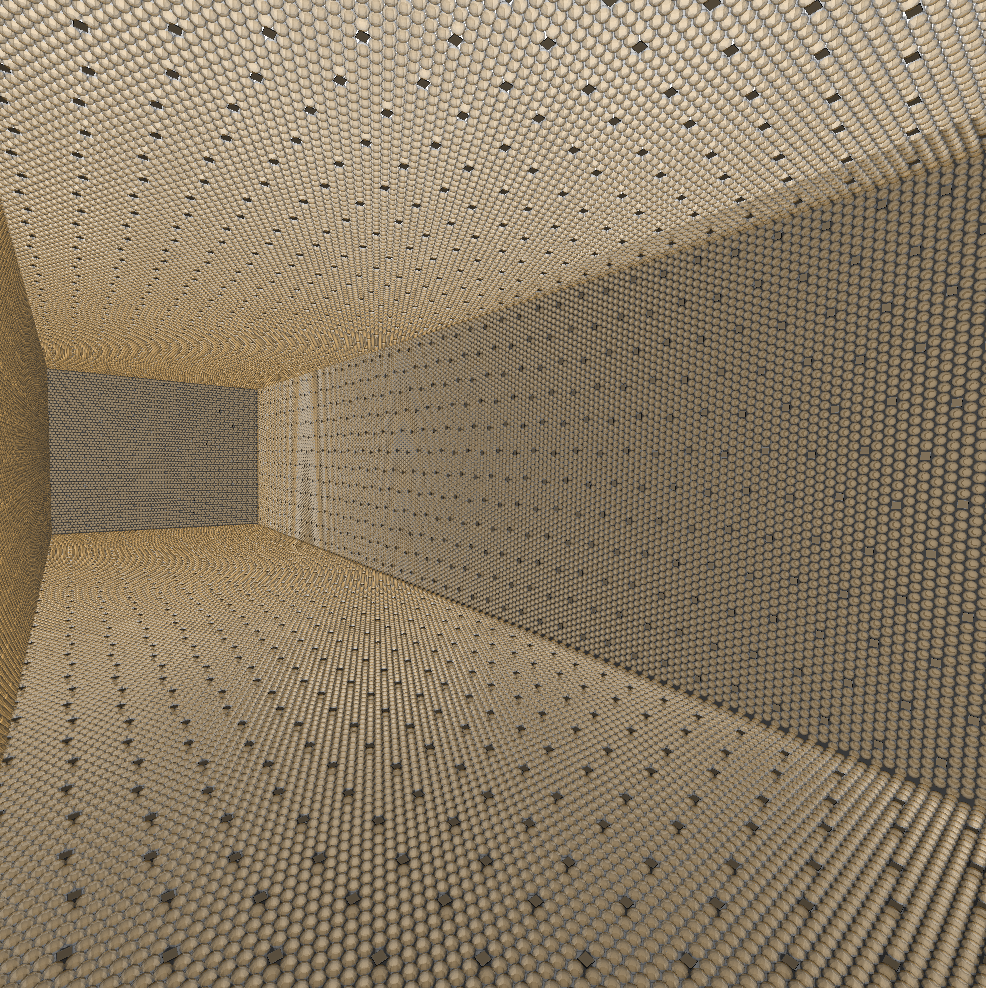}
\caption{The \theia detector.  
{\it Left  panel:} \theia-25 sited in the planned fourth DUNE cavern; 
{\it Right panel:} an interior view of \theia-25 modeled using the Chroma optical simulation package~\cite{chroma}. 
Taken from~\cite{theiawp}.}
\label{f:detector25}
\end{figure*}

\begin{figure*}[htp!]
\centering
\includegraphics[width=0.26\textwidth]{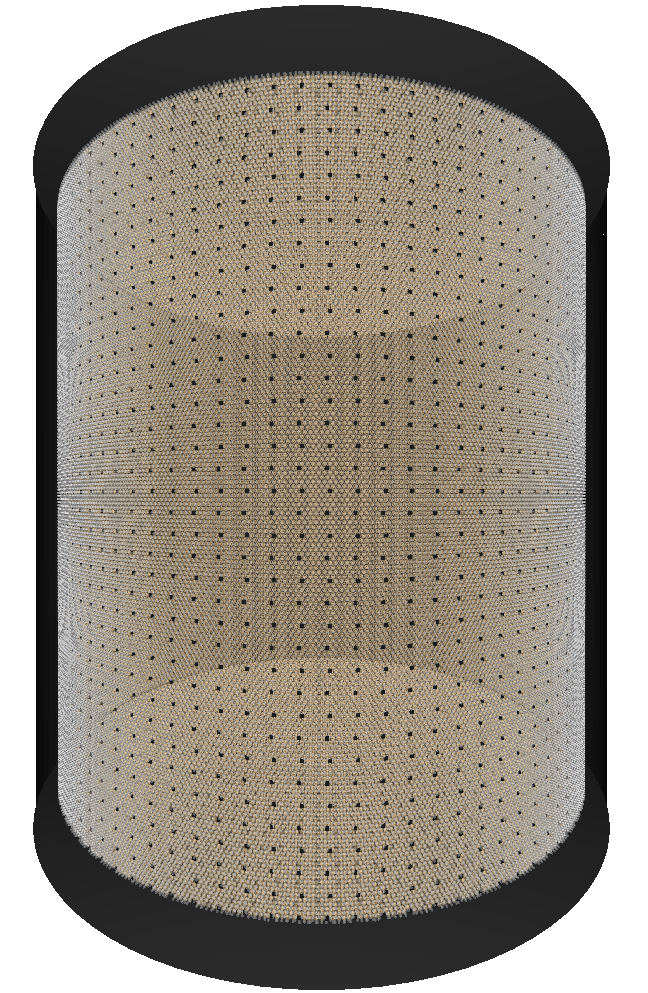}
\includegraphics[width=0.36\textwidth]{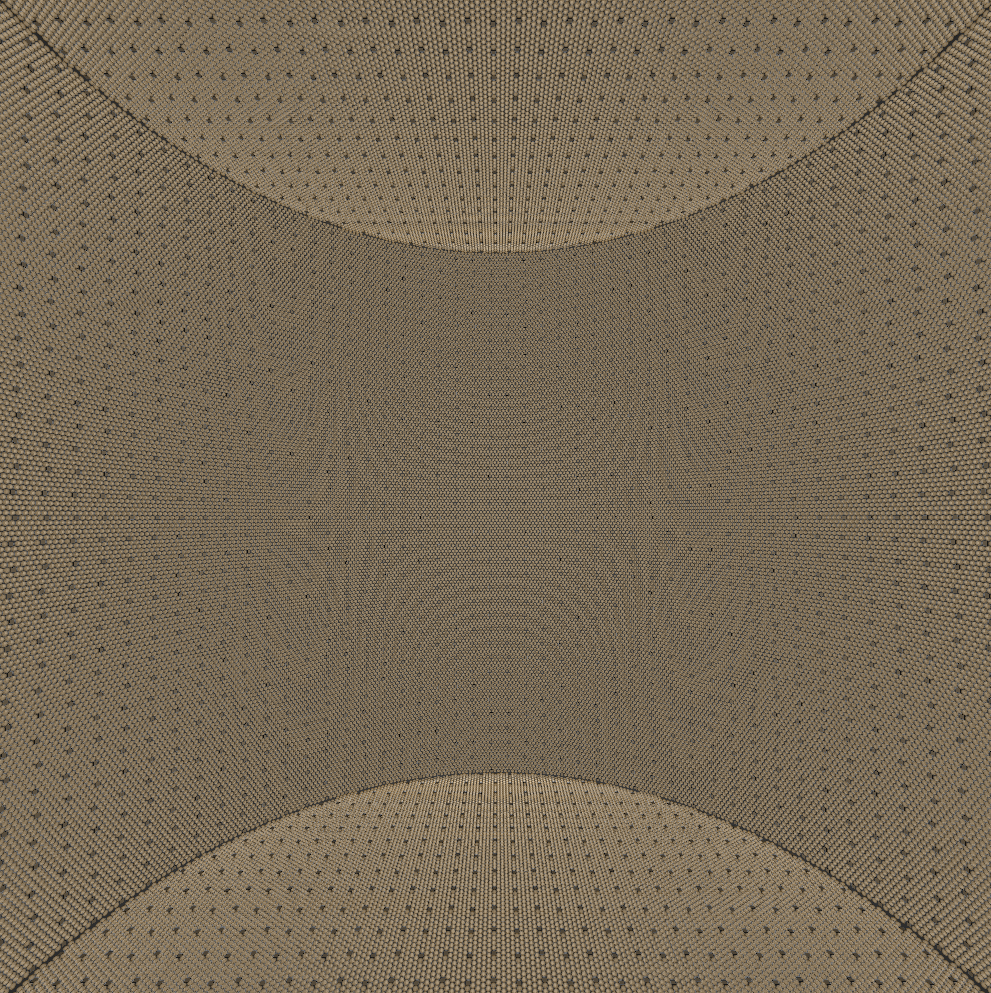}
\caption{The \theia detector.  
{\it Left panel:} Exterior view of \theia-100 in Chroma;
{\it Right panel:} an interior view of \theia-100 in Chroma.  In all cases, \theia has been modelled with 86\% coverage using standard 10-inch PMTs, and 4\% coverage with LAPPDs, uniformly distributed, for  illustrative purposes.  Taken from~\cite{theiawp}.}
\label{f:detector100}
\end{figure*}

We consider two scenarios, one in which \theia would reside in a cavern the size and shape of those planned for the Deep Underground Neutrino Experiment (DUNE), called \theia-25, and a larger 100-ktonne version (\theia-100) that could achieve an even broader and more sensitive scientific program (Figs.~\ref{f:detector25} and~\ref{f:detector100}).

Table~\ref{t:physics} summarizes the physics reach of \theia.  The full description of the analysis in each case can be found in~\cite{theiawp}. This broad program would be addressed using a phased approach, as shown in Table~\ref{tbl:phases}.

\begin{table*}[!h]
\centering
\caption{  \theia physics reach.  Exposure is listed in terms of the fiducial volume assumed for each analysis. 
For NLDBD the target mass assumed is the mass of the candidate isotope within the fiducial volume.
\label{t:physics}}
\begin{tabular}{l l l} 
\hline\noalign{\smallskip}
 {\bf Primary Physics Goal} & {\bf  Reach} & {\bf  Exposure / assumptions } \\
\noalign{\smallskip}\hline\noalign{\smallskip}
    Long-baseline oscillations 	& 	$> 5 \sigma$ for 30\% of $\delta_{CP}$ values	&	524 kt-MW-yr \\
   Supernova burst			&	$<1(2)^\circ$ pointing accuracy	& 100(25)-kt detector, 10~kpc	\\
 								& 20,000 (5,000) events  &\\
   DSNB 	&	$ 5 \sigma$ discovery	&	125 kton-yr \\
    CNO neutrino flux	&	$< $ 5 (10)\%	& 300 (62.5) kton-yr	\\
     Reactor neutrino detection	&	2000 events	&	100 kton-yr \\
     Geo neutrino detection 	&	2650 events	& 100 kton-yr	\\
     NLDBD               	&    	T$_{1/2} > 1.1\times10^{28}$~yr	 &     211 ton-yr $^{130}$Te	\\
     Nucleon decay $p\rightarrow \overline{\nu}K^{+}$ 	&	$T>3.80\times10^{34}$~yr (90\% CL)	& 800 kton-yr	\\
\noalign{\smallskip}\hline
\end{tabular}
\end{table*}

\begin{table*}
\centering
\caption{\theia physics goals and phased program.  Each successive phase adds to the breadth of the physics program.   The configuration column lists potential approaches to each phase, rather than a finalized detector design.
\label{tbl:phases}}
\begin{tabular}{c l l l l} 
\hline\noalign{\smallskip}
{\bf Phase} & {\bf Primary Physics Goals} & {\bf Detector capabilities} & {\bf Configuration options} \\
\noalign{\smallskip}\hline\noalign{\smallskip}
  I         &  Long-baseline oscillations 			& High-precision ring imaging & Low-yield WbLS  & \\
            &  $^{8}$B flux  					&&  Low photosensor coverage \\ 
            & Supernova burst, DSNB 			&& Fast timing     & \\ \hline
  II        & Long-baseline oscillations  			& Low threshold &  High-yield WbLS or slow LS & \\         
            &  $^8$B MSW transition    			& Cherenkov/scintillation  &  Potential $^7$Li loading  \\
            &  CNO, $pep$ solar 				&separation  & High photosensor coverage     \\
            &  Reactor and geo $\bar{\nu}$  		&High light yield&   Potential dichroicon deployment \\ 
            & Supernova burst ($\bar{\nu}_e$ and $\nu_e$), & &  & \\ 
            & DSNB ($\nu_e$ and $\bar{\nu}_e$) \\ \hline
 III        & $0\nu\beta\beta$  				& Low threshold&  Inner vessel with \\               
        & $^8$B MSW transition  				&Cherenkov/scintillation & LAB+PPO+isotope  \\  
            &  Reactor and geo $\bar{\nu}$  		&separation&  High photosensor coverage  \\ 
            &  Supernova burst and DSNB ($\bar{\nu}_e$) &High light yield& Potential dichroicon deployment\\ 
\noalign{\smallskip}\hline
\end{tabular}
\end{table*}

\subsection{Long baseline program}

Deployment of \theia in the broadband neutrino beam being built for the Long Baseline Neutrino Facility (LBNF)~\cite{lbnf1,lbnf2} would offer excellent sensitivity to neutrino oscillation parameters, including CP violation.  Advances in Cherenkov ring imaging techniques lead to improved particle identification and ring counting, which greatly improves background rejection.  
Improvements include a Boosted Decision Tree. 
Full details can be found in~\cite{theiawp}.
The fiTQun event reconstruction package, fully implemented in the most recent T2K analyses~\cite{Abe:2018}, was used for \theia sensitivity studies. This enhances the sensitivity to neutrino oscillations by: 
\begin{enumerate}
\item Improved particle identification using boosted decision trees removes 75\% of the neutral current background, relative to the previous analysis,
due to improvements in the detection of the faint second ring in boosted $\pi^{0}$ decays;
\item Improved electron/muon particle identification allows for an additional sample of 1-ring, one-Michel-electron events from $\nu_{e}$-CC$\pi^{+}$ interactions, without significant contamination from $\nu_{\mu}$ backgrounds
\item Multi-ring $\nu_{e}$ event samples can now be selected with sufficient purity to further enhance sensitivity to
neutrino oscillation parameters.
\end{enumerate}



\begin{figure*}[h!]
  \centering
  \includegraphics[width=0.45\textwidth]{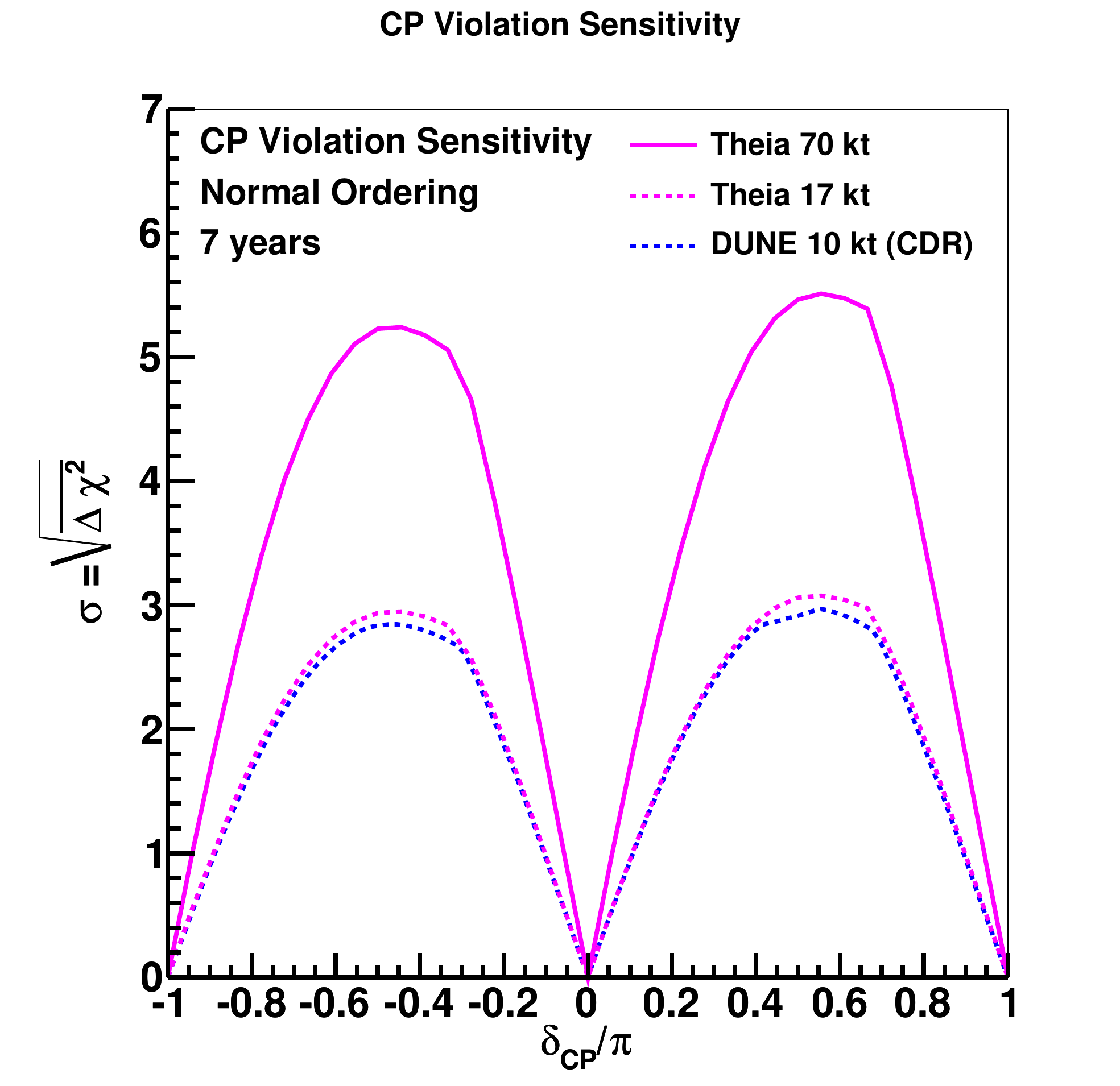}
  \includegraphics[width=0.45\textwidth]{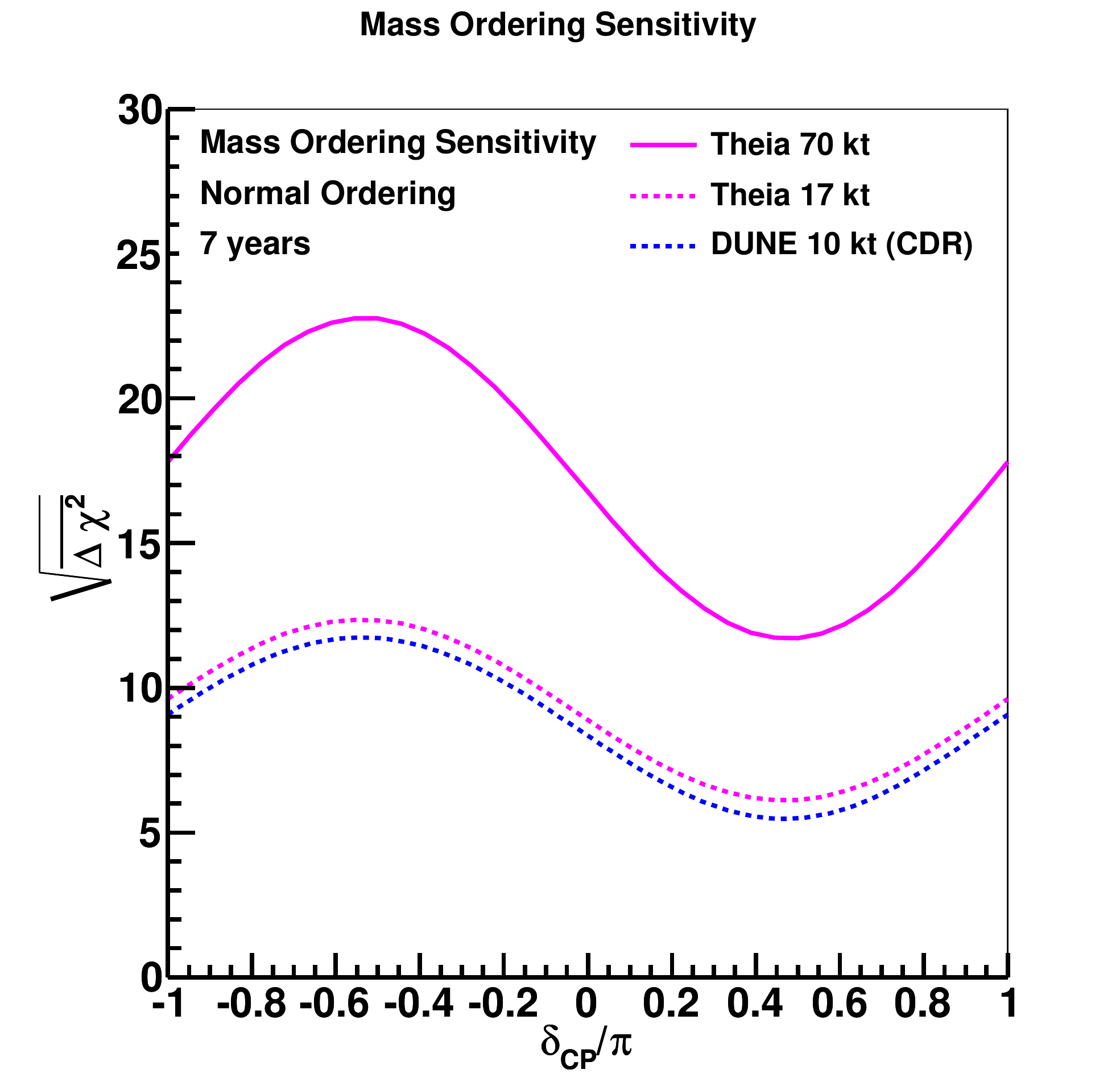}
  \caption{Sensitivity to CP violation (i.e.: determination that $\delta_{CP} \ne$ 0 or $\pi$) (left) and sensitivity
    to determination of the neutrino mass ordering (right), as a function of the true value of $\delta_{CP}$, for the \theia 70-kt fiducial volume detector (pink). Also shown are sensitivity curves for a 10-kt (fiducial) LArTPC (blue dashed) compared to a 17-kt (fiducial) \theia-25 WCD detector (pink dashed). Seven years of exposure to the LBNF beam
    with equal running in neutrino and antineutrino mode is assumed. LArTPC sensitivity is based on
    detector performance described by \cite{Alion:2016uaj}.}
  \label{fig:sens_lbl}
\end{figure*}

The sensitivity to long-baseline physics at \theia is seen to be comparable to the sensitivity of an equivalently-sized DUNE module~\cite{duneidr1,duneidr2,duneidr3}, as shown in Fig.~\ref{fig:sens_lbl}. The ability to measure long-baseline neutrino oscillations with a distinct set of detector systematic uncertainties and neutrino interaction uncertainties relative to the liquid argon detectors, would provide an important independent cross-check of the extracted oscillation parameter values.

\subsection{Solar neutrinos}

\begin{figure}[!t]
	\includegraphics[width=0.65\textwidth]{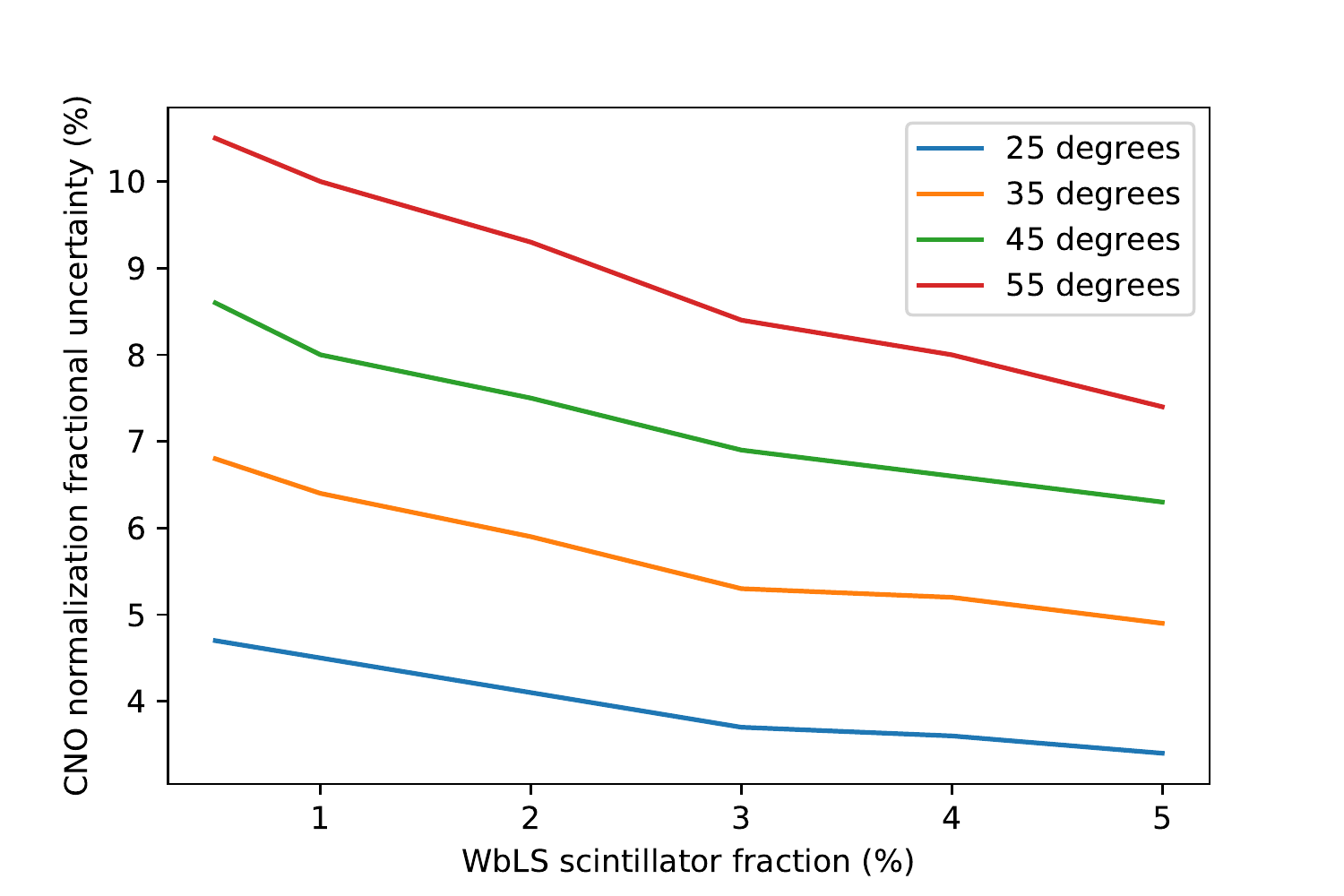}
	\caption{Fractional uncertainty on the CNO normalization parameter as a function of angular resolution and WbLS scintillator fraction, for \theia-100.} 
	\label{fig:tableplot}
\end{figure}

\theia will be sensitive to a precision measurement of neutrinos from the Sun's Carbon-Nitrogen-Oxygen cycle (Fig.~\ref{fig:tableplot}).  Although observed by Borexino~\cite{2020Natur.587..577B}, measurements to-date lack sufficient precision to distinguish between predictions from different solar models~\cite{Bahcall:2005va}.  Much of the power of a measurement at \theia will come from the directional sensitivity enabled by the combination of novel technology planned, coupled with high statistics from the large target volume.
\theia will also provide a high-statistics, low-threshold (MeV-scale)
measurement of the shape of the $^8$B solar neutrinos and thus search for new
physics in the MSW-vacuum transition region~\cite{friedland2004,minakata2012}. 
Other exciting physics opportunities related to solar neutrinos include tests of solar luminosity through precision measurements of $pep$ and $pp$ neutrinos; tests of the solar temperature; and, potentially, separation of the different components of the CNO flux to probe the extent to which this cycle is in equilibrium in the Sun's core~\cite{HRS}.  

\subsection{Supernova neutrinos}

Should a supernova occur during
\theia  operations, a high-statistics detection of the $\bar{\nu}_e$ flux will be
made -- literally complementary to the  detection of the $\nu_e$ flux in the DUNE liquid argon
detectors. At 100\,kt, \theia will more than double the statistics expected for both SK and JUNO combined in $\bar\nu_e$-induced IBD signals and add hundreds of events for $\nu_e$'s and $\nu_x$'s (Tab.~\ref{tab:snrates}). Detection in WbLS will be flavor-resolved since delayed signals (neutrons from IBD, ${^{16}{\rm N}}$) and de-excitation gamma rays from NC reactions on oxygen can be clearly identified (Fig.~\ref{fig:snspectra}). Together with a good energy resolution, this will provide great sensitivity to look for energy-dependent oscillation patterns, e.g.~the spectral swaps induced by collective oscillations, resonant flavor conversion during the neutronization burst (where \theia will provide meaningful statistics of $\nu_e$-electron elastic scattering) and complementary measurements of Earth matter effects (DUNE providing a $\nu_e$ signal and JUNO and SK $\bar\nu_e$ signal from the other side of the planet).

\theia (especially at 100\,kt) will be a great detector for complementary measurements with other neutrino experiments and multi-messenger observations involving electromagnetic or gravitational waves. The presence of a clear neutron-tag for IBD events means that SN pointing accuracy can reach sub-degree accuracy for follow-up astronomical observations. High statistics and energy resolution will permit to resolve time-dependent spectral features that can be correlated to other observations, e.g.~with gravitational wave emission in the early accretion phase (SASI) \cite{Kopke:2017req}. Finally, \theia will feature excellent sensitivity to pre-Supernova neutrinos, permitting a $3\sigma$ detection to an unprecedented distance of up to 3~kpc.

\subsection{Diffuse supernova neutrino background}

With a very deep location and with the detection of a combination of scintillation and Cherenkov light~\cite{Nakamura:2016kkl,branchesi2016}, \theia will have world-leading sensitivity to make a detection of the Diffuse Supernova Neutrino Background (DSNB) antineutrino flux~\cite{SK_DSNB:2015,priya2017,Sawatzki:2019}. 
With a target mass several times the size of SK or JUNO, \theia-100 will obtain a $\sim$5$\sigma$ discovery of the DSNB in less than 1 year of data taking and reach ${\cal O}(10^2)$ DSNB events within $\sim$5 years. Even the smaller \theia-25 will profit considerably from the dual detection of Cherenkov and scintillation signals that offer a background discrimination capability unparalleled by Gd-doped water or pure organic scintillator: For instance, a signal efficiency of $>$80\,\% can be maintained while reducing the most crucial background from atmospheric neutrino NC interactions to a residual $\sim$1.3\,\% \cite{Sawatzki:2019}.

\begin{figure}[htp!]
\centering
\includegraphics[width=0.9\textwidth]{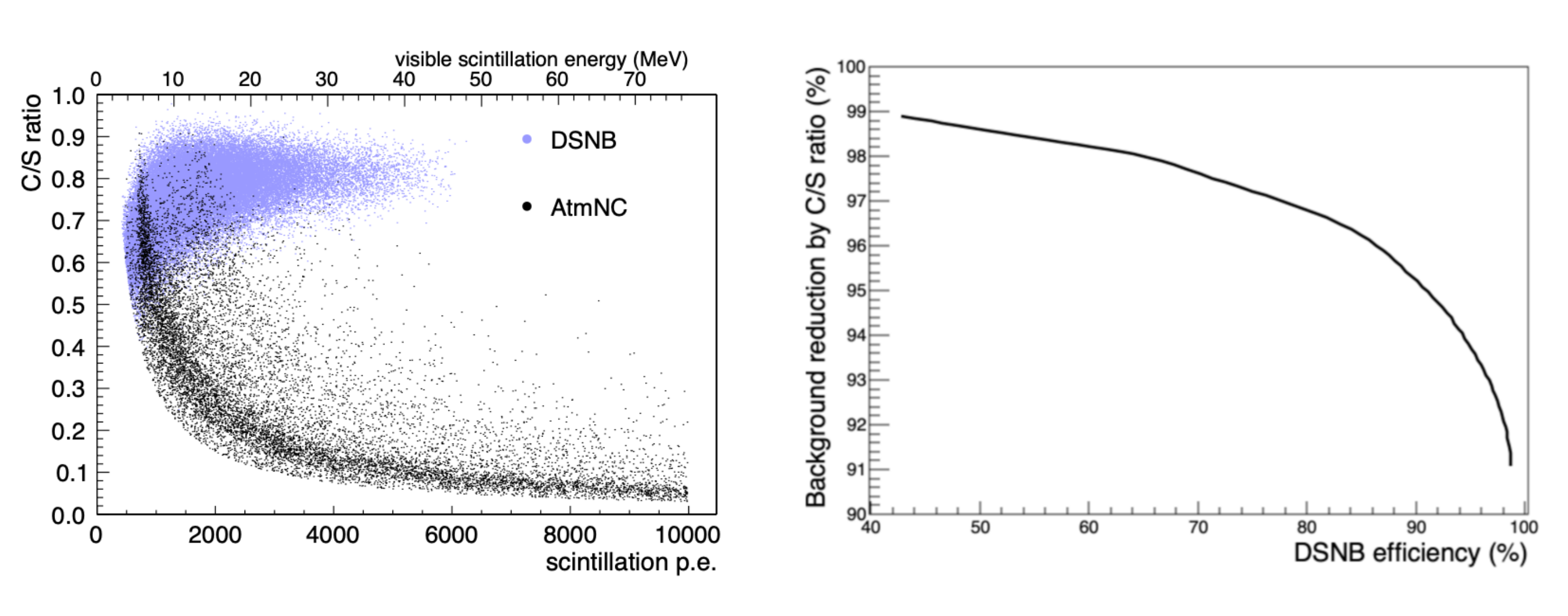}
\caption{{\it Left panel:} The C/S ratio offers a powerful tool for the discrimination of positron-like DSNB ({\it blue}) and hadronic prompt events from atm-NC reactions ({\it black}). While most background events feature no Cherenkov light and, as such, a C/S ratio of 0, some final-state $\gamma$ rays result in a curved band of atm-NC events that leaks slightly into the signal region. {\it Right panel:} atm-NC background reduction as function of the DSNB signal efficiency.}
\label{fig:dsnb_csratio}
\end{figure}

The main virtue of \theia lies with the excellent background discrimination capabilities of the hybrid detector. 
While $e^\pm$ signals feature a high Cherenkov-to-scintillation (C/S) ratio, that of non-relativistic nuclear recoils is practically zero. The left panel of Fig.~\ref{fig:dsnb_csratio} shows the C/S ratios of signal and atm-NC events as a function of visible energy, demonstrating clear separation. The right panel  shows the relation between signal efficiency and background reduction factor \cite{Sawatzki:2019}.

\theia-25 will require about 6\,years of data taking to achieve a $5\sigma$ discovery of the DSNB signal (assuming standard predictions for flux and spectral energy) \cite{Sawatzki:2019}. Combined with SK+Gd and JUNO, the three detectors will acquire about 5-10 DSNB events per year (with $\sim$40\% of statistics from \theia), so that a spectroscopic analysis of the DSNB based on ${\cal O}(10^2)$ events will become feasible over 10-20 years. At the same time, the C/S signatures of the large atm-NC event sample recorded in WbLS will enable an in-depth study of this most relevant background and will help to reduce the corresponding systematic uncertainties as well for SK-Gd and especially JUNO.

\subsection{Geological and reactor neutrino measurements}

Antineutrinos produced in the
crust and mantle of the Earth will be measured precisely by \theia with statistical uncertainty far exceeding all detectors to date. 
The rate and energy spectrum of global antineutrino interactions varies dramatically with surface location \cite{agm15}. 
Observations of Earth antineutrinos, or geo-neutrinos, probe the quantities and distributions of terrestrial heat-producing elements uranium and thorium. The quantities of these elements gauge global radiogenic power, offering insights into the origin and thermal history of Earth \cite{dye12}. 

The predicted rate of geo-neutrino interactions per kT-year at SURF is $26.5$ ($20.7$ U and $5.8$ Th), which corresponds to a flux of $4.90 \pm 0.13 \times 10^6$ cm$^{-2}$ s$^{-1}$, assuming perfect background suppression, Th/U $=3.9$, and statistical uncertainty only. 
A measurement at the predicted SURF rate would be almost $2 \sigma$ greater than the KamLAND measurement after an exposure of $50$ kT-y. This would provide the first evidence for surface variation of the geo-neutrino flux. With thousands of geo-neutrino events \theia would precisely measure the uranium and thorium components of the energy spectrum with the potential to test models of differential partitioning and transport of these trace elements between silicate mantle and crust types.

The expected rate of reactor antineutrino interactions at SURF is $\sim 20$ per kT-year.  A detailed study of the antineutrino capabilities of \theia is in preparation for a separate publication.

\subsection{Neutrinoless double beta decay}

The \theia search for neutrinoless double beta decay (NLDBD) aims for
sensitivity to the non-degenerate normal hierarchy parameter space
within the canonical framework of light Majorana neutrino exchange and
three-neutrino mixing, at the level of $m_{\beta\beta}\sim5$ meV.
This is achieved through the loading of a very large
mass of a NLDBD candidate isotope into an ultra-pure LS
target, held inside a balloon or thin vessel within the larger WbLS detector. The main advantages of this technique are the higher light yield of pure LS compared to WbLS, the higher radiopurity, and  reduced contamination from external backgrounds. Coincidence and topological particle identification techniques help to further reduce background.

Two cases are considered: loading with 3\% enriched Xenon (89.5\% in $^{136}$Xe) and 5\% natural Tellurium (34.1\% in $^{130}$Te), each loaded into a high light yield LS held inside an 8-m radius containment balloon deployed within \theia-100. All known significant sources of background are considered, including cosmogenics, internal radioactivity, 2$\nu\beta\beta$, solar neutrinos, and external sources of radioactivity.

The sensitivity is estimated via a single-bin counting analysis. Since several backgrounds do not scale with isotope mass,  Monte Carlo is used to evaluate the background expectation, establish a confidence region using the Feldman-Cousins frequentist approach, and derive an expected limit on the NLDBD half-life.



The expected sensitivity (90\% CL) for 10 years of data taking, using phase space factors from \cite{2012PhRvC..85c4316K} and matrix element from \cite{Barea:2013wb} (g$_{A}$=1.269) is:
\begin{eqnarray*}
\mathrm{\bf Te:}~~
  T_{1/2}^{0\nu\beta\beta} > 1.1\times10^{28}~\mathrm{y},~
  m_{\beta\beta} < 6.3~\mathrm{meV}\\
  \mathrm{\bf Xe:}~~
  T_{1/2}^{0\nu\beta\beta} > 2.0\times10^{28}~\mathrm{y},~
  m_{\beta\beta} < 5.6~\mathrm{meV}
\end{eqnarray*}

\begin{figure}
\centering
 \includegraphics[width=0.65\textwidth]{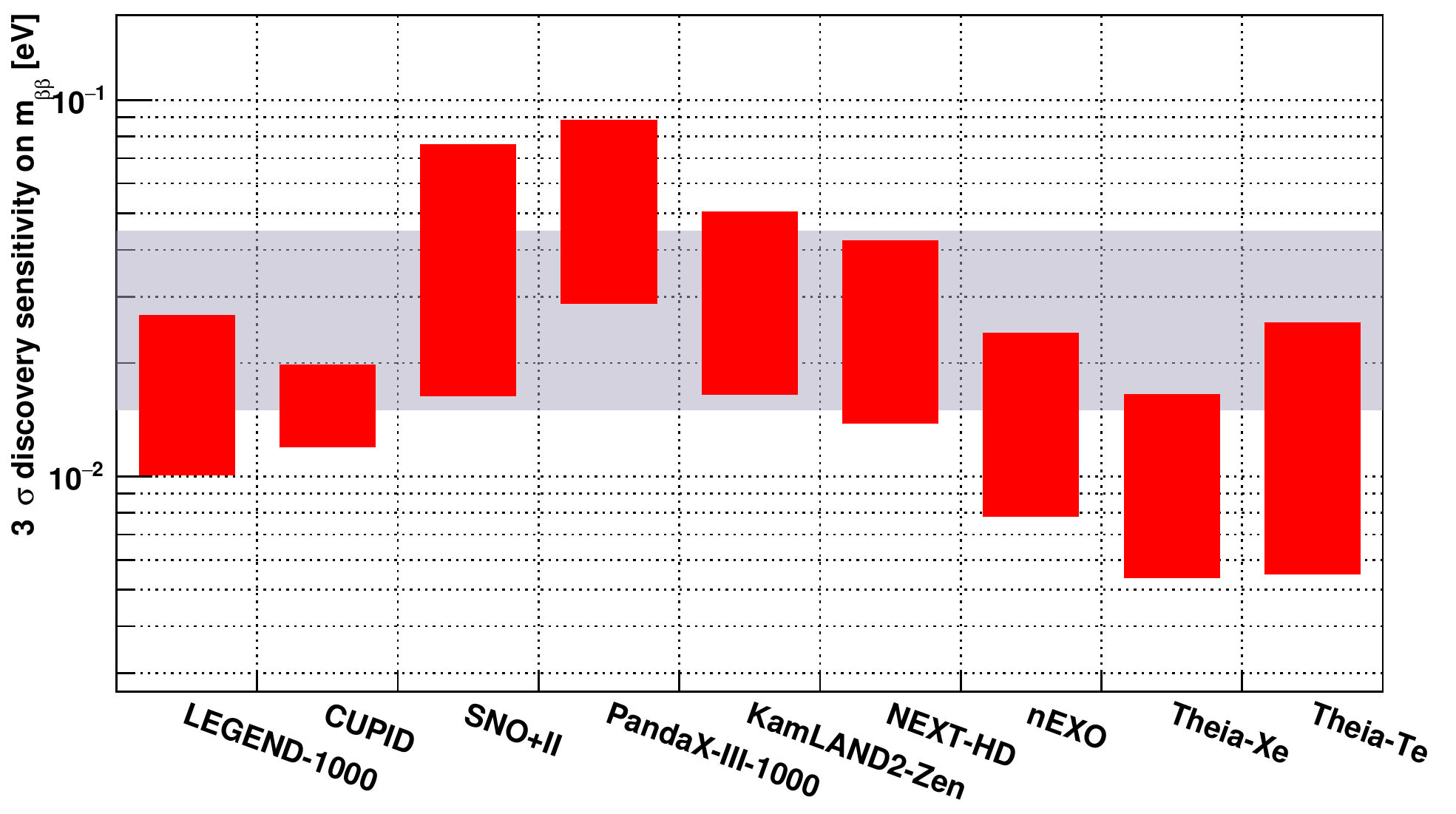}
 \caption{Discovery sensitivity (3$\sigma$) for  proposed future  experiments. The grey shaded region corresponds to the parameter region allowed in the Inverted Hierarchy of the neutrino mass. The red error bars show the $m_{\beta\beta}$ values such that an experiment can make at least a 3$\sigma$ discovery,
within the range of the nuclear matrix elements for a given isotope. The parameters of the other experiments are taken from Refs.~\cite{Agostini:2017jim,Gomez-Cadenas:2019sfa,Galan:2019ake,CUPIDInterestGroup:2019inu,Giuliani:2019uno}.}
\label{fig:dbdcomp}
\end{figure}

A comparison of this sensitivity to other experiments is shown in Fig.~\ref{fig:dbdcomp}.



\subsection{Nucleon decay}

\begin{figure}[h]
    \centering
    \includegraphics[width=0.48\textwidth]{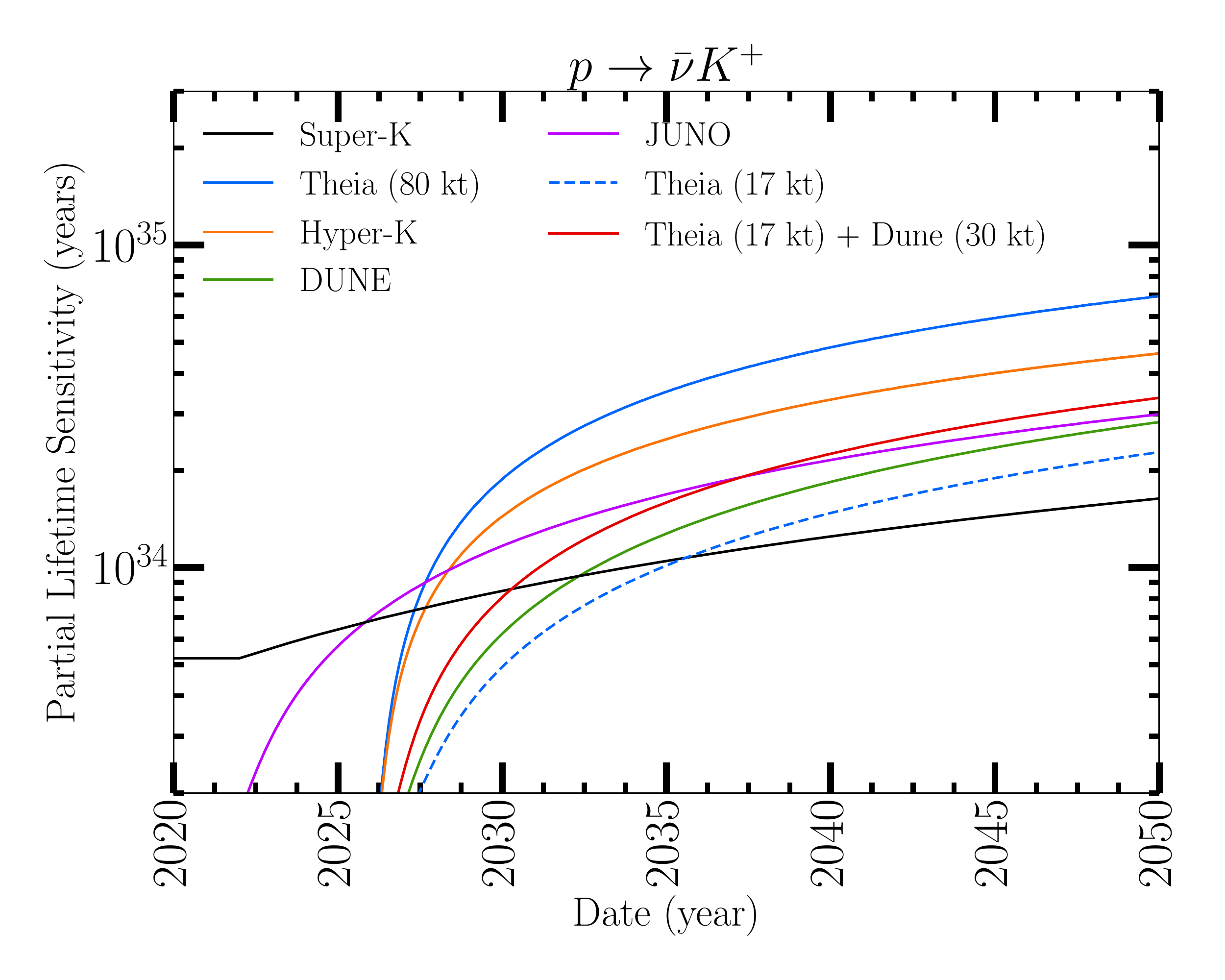}
    \includegraphics[width=0.48\textwidth]{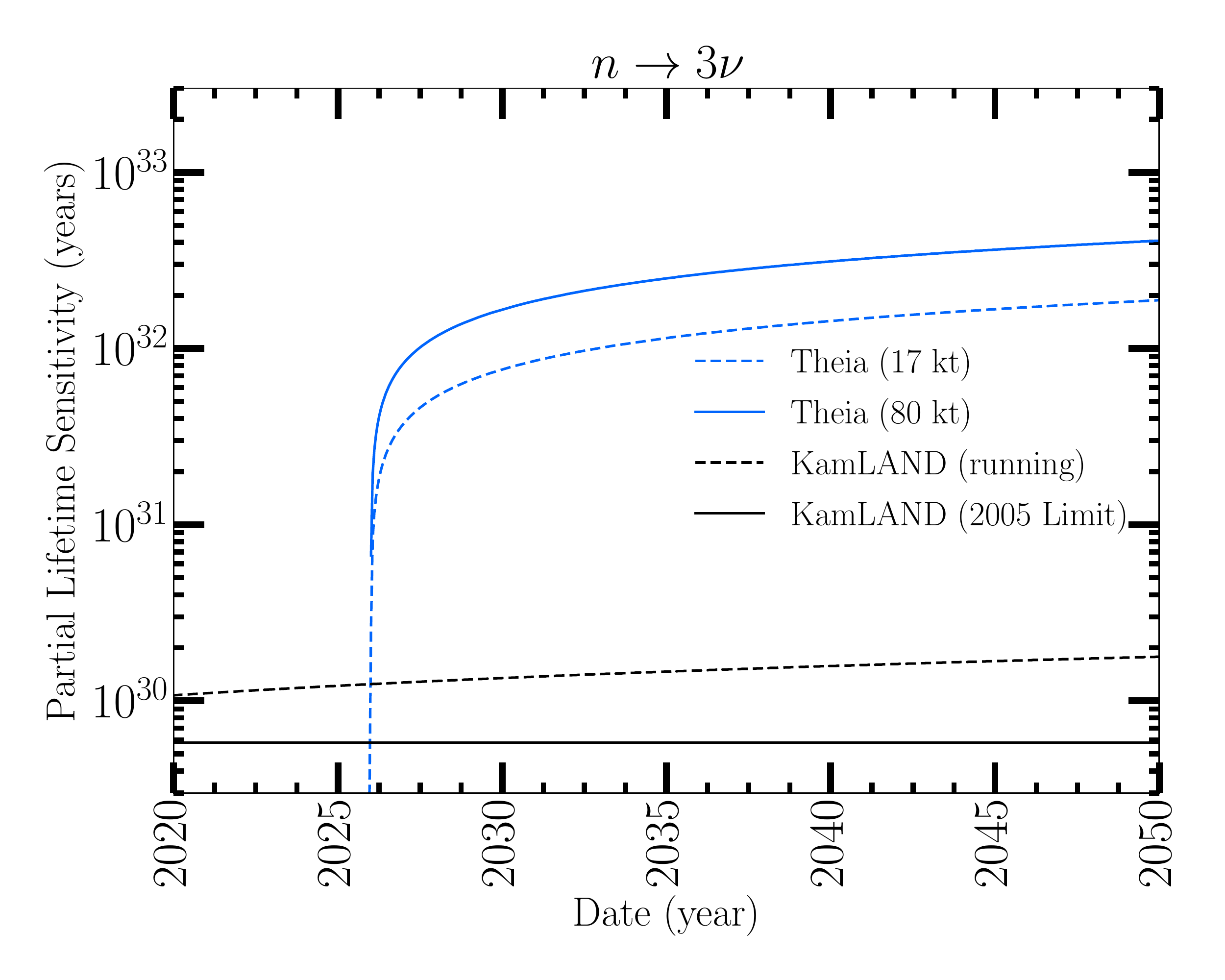}
    \caption{(Left) Sensitivity for $p \rightarrow \bar{\nu}K^+$ is highest for \theia, closely followed by the
    Hyper-K detector, whereas JUNO and DUNE will perform similarly. (Right) The large size and depth of \theia means that other next generation
    detectors are unlikely to be competitive when looking for very low energy
    modes of nucleon decay.
    }
    \label{fig:nuk_sensitivity}
\end{figure}

The \theia detector has the size and resolution to
contribute to searches for nucleon decay and, in certain modes, provide the dominant
experimental measurement. In the case of modes like $p\rightarrow e^{+}\pi^0$, the efficiency of \theia would be similar to current detectors like Super-Kamiokande and future detectors like Hyper-Kamiokande; thus, \theia-25 would add only marginally to the DUNE program, and \theia-100 would add in proportion to the exposure. For modes like $p\rightarrow \overline{\nu} K^{+}$ the contribution to the global sensitivity would be significant, and for modes like $n\rightarrow 3\nu$ \theia would be world-leading. These are discussed in more detail in~\cite{theiawp}.  Examples of \theia sensitivity are shown in Fig.~\ref{fig:nuk_sensitivity}.

\subsection{Sterile neutrinos}

The 100 kton (25 kton)  \theia detector  represents a particularly promising venue for a decisive test of sterile neutrino hypotheses using neutrino and antineutrino generators,  to observe a \textit{distance-dependent} neutrino flux from the source at the distances of the order of oscillation length. 

In case of sterile neutrino $\Delta m^2 > 1$ eV$^2$, the oscillation distance  is of the order of couple of meters. \theia spans over tens of meters of distance and the source can be placed within meters of the target, creating baseline comparable to the sterile neutrino oscillation length.  Observation of the oscillation pattern in the distant-dependent measurement in WbLS/LS  would represent the most convincing proof of the existence of sterile neutrinos and their oscillation with the other three flavors.

\theia has high potential to carry out a definitive search for sterile neutrinos free of statistical limitations that plague current generation of experiments, in the high sterile neutrino mass regime of $\Delta m^2 > 1$ eV$^2$. With its large target mass of up to 100 kton,  a tremendous number of IBD interactions from $^{144}$Ce-$^{144}$Pr source  could be collected allowing a detailed, high statistics study of the position dependent neutrino flux. \theia could also directly cross-check a high-confidence electron neutrino measurement from GA and BEST  with an independent detection method.

\theia is uniquely suited to perform both electron neutrino and electron antineutrino disappearance measurement. As a result, a robust picture of sterile neutrino landscape in the disappearance channel can emerge from a single detector.

\section{Conclusions}

The \theia detector design represents an important step forward in the realization of a new kind of large optical detector: a hybrid Cherenkov/scintillation detector. Such an instrument is made possible by the convergence of three technological breakthroughs: (1) the development of ultra-fast and/or chromatically sensitive photosensors, (2) the ability to make novel target materials that produce detectable levels of both kinds of light, and (3) highly sophisticated pattern recognition and data analysis techniques that have moved well beyond the relatively simple methods of a decade ago. In addition, the coming availability of the new Long Baseline Neutrino Facility enables a broad program due to the deep depth and powerful neutrino beam available at the site.

This paper has detailed the exciting possibilities of \theia for new scientific discovery across a broad spectrum of physics, including long baseline neutrino measurements of oscillations and CP violation searches, solar neutrino measurements of unprecedented precision and scope, and the potential to extend the reach of neutrinoless double beta decay searches to the mass scales implied by a Normal Ordering of neutrino masses. In addition, \theia represents a major step forward for advancement in other fields -- ranging from detection of the Diffuse Supernova Background flux to a measurement of geo-neutrinos with unprecedented statistics. More detail on the detector and physics program can be found in~\cite{theiawp}.

\begin{table}[h!]
\begin{minipage}[b]{0.4\textwidth}
\begin{tabular}{llr}
\hline
\multicolumn{2}{l}{Reaction} & Rate \\
\hline
(IBD) & $\bar\nu_e+p\to n+e^+$ & 19,800 \\
(ES) & $\nu+e \to e+\nu$ & 960 \\
($\nu_e$O) & ${^{16}\rm{O}}(\nu_e,e^-){^{16}{\rm F}}$ & 340 \\
($\bar\nu_e$O) & ${^{16}\rm{O}}(\bar\nu_e,e^+){^{16}{\rm N}}$ & 440 \\
(NCO) & ${^{16}\rm{O}}(\nu,\nu){^{16}{\rm O}^*}$  & 1,100 \\
\hline
\end{tabular}
\caption{Event rates expected in \theia-100 for an SN at 10~kpc distance (GVKM model \cite{Gava:2009pj} and SNOwGLoBES), incl.~Inverse Beta Decays (IBDs), elastic scattering off electrons (ES), charged-current ($\nu_e$O,$\bar\nu_e$O) and neutral-current (NCO) interactions on oxygen.}
\label{tab:snrates}
\end{minipage}\hfill
\begin{minipage}[b]{0.55\textwidth}
\centering
\includegraphics[width=\textwidth]{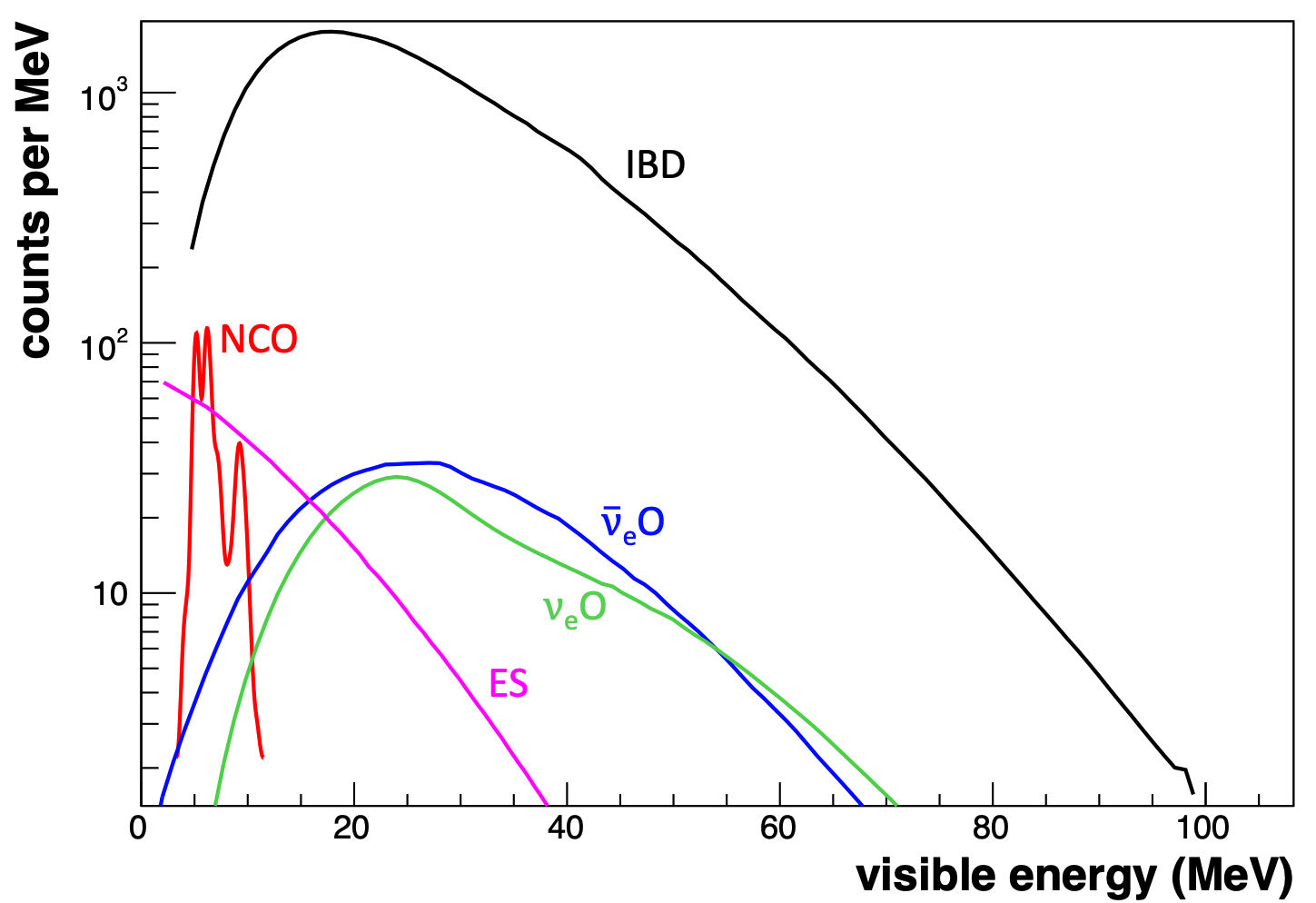}
\captionof{figure}{Visible energy spectra of the prompt events, corresponding to the event rates of Tab.~\ref{tab:snrates} with a Gaussian energy resolution of 7\,\% at 1\,MeV.}
\label{fig:snspectra}
\end{minipage}
\end{table}


\clearpage

\section*{Acknowledgments}

This material is based upon work supported by: the U.S. Department of Energy Office of Science under contract numbers DE-AC02-05CH11231, DE- SC0009999, DE-FG02-88ER40893, and DE-FG02-88ER40479; the U.S. Department of Energy National Nuclear Security Administration through the Nuclear Science and Security Consortium under contract number DE-NA0003180; the U.S. National Science Foundation award numbers 1554875 and 1806440; Funda{\c c}{\~ a}o para a Ci{\^ e}ncia e a Tecnologia (FCT-Portugal); and the Universities of California at Berkeley and Davis.


\renewcommand{\bibname}{References}
\bibliographystyle{utphys}
\bibliography{Theia}

\end{document}